\documentclass[11pt,twoside]{article}

\usepackage{asp2006}
\usepackage{epsf}
\usepackage{psfig}
\usepackage{lscape}
\usepackage{graphicx}

\begin{document}
\title{The period-luminosity relation of Mira variables
in NGC 6388 and NGC 6441}
\author{Noriyuki Matsunaga \altaffilmark{1} and the IRSF/SIRIUS team} 
\affil{Institute of Astronomy, School of Science, the University of Tokyo,
2-21-1 Osawa, Mitaka, Tokyo 181-0015, Japan}
\altaffiltext{1}{matsunaga@ioa.s.u-tokyo.ac.jp}

\begin{abstract} 
We report a result of our near-infrared repeated observations of
Mira variables in the globular clusters NGC 6388 and 6441.
These two clusters are known as peculiar clusters with
blue horizontal branch stars and RR Lyr variables which are
unexpected for their relatively high metallicities.
We derive their distance moduli by fitting the Mira variables
in the period-luminosity relation. This is the first distance estimates,
for these clusters, which is observationally obtained
in an independent way from the horizontal branch stars. 
The obtained distances revealed that the absolute magnitudes
of the peculiar RR Lyr variables are similar to
the metal-poor ones of [Fe/H]$=-2$ dex.
It is suggested that the constraint we found should be reproduced
by any theories to explain the horizontal morphology
of these peculiar clusters.
\end{abstract}



\section{Introduction}

The globular clusters NGC 6388 and 6441 have attracted many researchers
for their peculiar populations of horizontal branch (HB) stars in this decade.
In spite of their relatively high abundances ([Fe/H]$\sim -0.6$),
they have extended blue HB stars and RR~Lyr variables
(\citealt{Rich-1997}; \citealt{Layden-1999}). Furthermore,
the red HB slopes upward towards the blue.
This feature is quite different from that of the usual HB
of globular clusters, and discovered in these clusters
for the first time. No canonical theory can reproduce this feature
\citep{Sweigart-1998}. It was discovered that there is (are)
non-canonical effect(s) at work (e.g. helium enrichment or rotation),
but no final conclusion has yet been reached.

It is important to place and see this interesting feature on
the absolute magnitude scale. Different non-canonical process may have
different effects on the luminosity. Horizontal branch stars and
RR Lyr variables themselves are distance indicators which are
most frequently used for globular clusters. However, it is unknown
whether the traditional measure is applicable to the peculiar objects.
In order to derive their absolute magnitudes, it is necessary to derive
distances of the clusters in an independent way.
Mira variable can be a good distance indicator especially for
relatively metal-rich population (see \citealt{Feast-2004}, for the review).
In this contribution, we will utilise Mira variables for the distance
estimation, and discuss its impact on the peculiar HB.

\section{Observation and analysis}

We are conducting a near-infrared survey by using the Infrared Survey Facility
(IRSF) and the Simultaneous three-colour Infrared Imager for Unbiased Survey
(SIRIUS) sited at SAAO, Sutherland, South Africa
(see \citealt{Matsunaga-2006a}, for some preliminary results).
From the monitoring survey, started in the spring of 2002, we obtained 
an extensive dataset for Mira variables in globular clusters
including NGC 6388 and 6441. The frequency of the observations for each cluster
is about once a month or more during April--August every year.
Many globular clusters have been observed more than 20 times in total.
NGC 6388 and 6441 were observed 31 and 40 times, respectively, excluding
some unused data with bad conditions (weather or seeing).

The analysis of Mira variables was conducted as follows.
At first, 10 or 15 dithered images were combined into one scientific image
by using the pipeline software for the standard data reduction 
(Nakajima, private communication). Photometry was executed by using
the DoPHOT software, and a time series of the results were compared
 with a reference list obtained for the best image for each cluster.
The magnitudes were standardised based on those of cluster giants
listed in the 2 Micron All-sky Survey (2MASS) point source catalogue.
The detail of the procedures was presented in \citet{Matsunaga-2006b}.

\section{Result}

\subsection{Mira variables in NGC 6388 and 6441}

\begin{figure}[!ht]
\begin{center}
\includegraphics[clip,width=0.7\hsize]{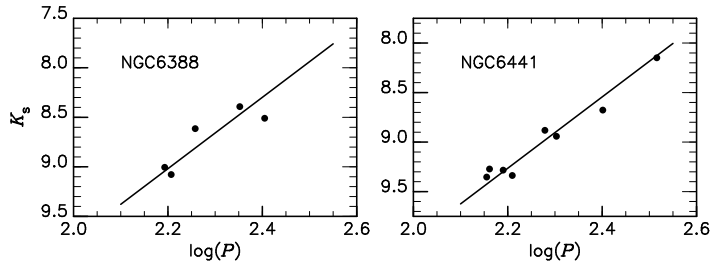}
\caption{Period-luminosity relation of Mira variables in NGC 6388 and 6441.
\label{fig:PLR}}
\end{center}
\end{figure}

We discovered one and four new Mira variables in NGC 6388 and 6441,
respectively. They are named NGC 6388 V1 and NGC 6441 V151--V154
after the list of newly reported variables in \citet{Corwin-2006}.
With all the known Mira variables, we obtained periods and
mean magnitudes for five and nine objects in the two clusters.
For most of them, such quantities are obtained for the first time.
The periods were found with the phase dispersion minimization (PDM) method
and the mean magnitudes were adopted by fitting sine curves to light curves.
The data will appear in a future paper.
The relation between the periods and the mean $K_{\rm s}$ magnitudes
was plotted in Figure \ref{fig:PLR}. NGC 6441 V151 was not plotted,
but is much brighter than expected from the relation. 
We conclude that this star is not a member of the cluster.
The slope of the linear relations overplotted was obtained
from the relation of Mira variables in the Large Magellanic Cloud (LMC),
as will be described below.

\subsection{Distance and the absolute magnitude of horizontal branch}

Now, we will derive the distance moduli of the clusters
from the period-luminosity relation of the Mira variables.
We use the relation of Mira variables in the LMC as a reference one.
From the data published by \cite{Ita-2004b}, we extracted
139 oxygen-rich Mira variables in the LMC with the conditions,
$100 \leq P \leq 350$ (days) and $J-K_{\rm s} \leq 1.4$ (mag).
Those with the long period ($P>350$) were excluded
because they are considered to be hot bottom burning stars
which are brighter than the period-luminosity relation \citep{Whitelock-2003}.
The colour condition is to take away carbon-rich Mira variables and
those with thick circumstellar shell.
The derived relation for the oxygen-rich objects is,
\begin{equation}
M_{K_{\rm s}} = -3.60 \log P + 1.36,
\label{eq:PLR}
\end{equation}
with the LMC distance modulus of 18.50 mag assumed.

Now, we compare the above relation with the relations in Figure \ref{fig:PLR}.
After the reddening correction for each cluster,
the distance modulus was obtained and listed in Table \ref{tab:result}.
The reddening values $E_{B-V}$ were adopted from
the catalogue by \citet{Harris-1996}, and we used the extinction law
of $A_{K_{\rm s}}/E _{B-V} = 0.365$ and $R_{V}=3.1$.
We also derived distance moduli for 47 Tuc and NGC 6637,
which have similar metallicities to NGC 6388 and 6441.
Although the period-luminosity relation may have a metallicity dependence
as discussed by \citet{Ita-2004a}, we can ignore it because
the four clusters have similar [Fe/H] values. The listed error sizes
include random scatters around the period-luminosity relation and
the errors of the reddening which are assumed to be 20 \% of $E_{B-V}$.
There is a systematic error from the assumed distance modulus of the LMC,
but it has no effect on the relative difference between the four clusters.

\begin{figure}[!ht]
\begin{center}
\includegraphics[clip,width=0.9\hsize]{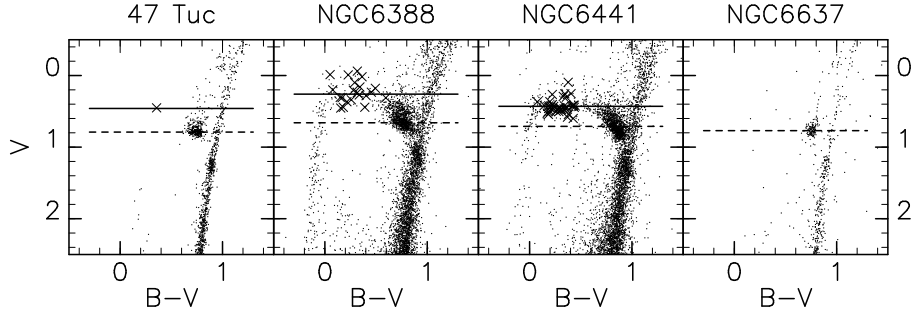}
\caption{Optical colour-magnitude diagrams for four clusters.
Data from \citet{Piotto-2002} were shifted to the absolute scale
with the distance moduli obtained in this contribution.
RR Lyr variables are plotted as crosses.
\label{fig:CMD}}
\end{center}
\end{figure}

By using the obtained distance moduli, we plotted their optical
colour-magnitude diagrams (CMDs) on the absolute scale (Figure \ref{fig:CMD}).
The optical data were from \citet{Piotto-2002}. We derived the absolute
magnitudes red HB stars, $M_V$(RHB), and those of RR Lyr
variables, $M_V$(\rm RR). The former was defined as a peak of
the distribution in the $(B-V, V)$ diagram. The apparent mean magnitudes of
RR Lyr variables were taken from \citet{Pritzl-2002} and \citet{Pritzl-2003}.
There is one RR Lyr variable found in 47 Tuc, and we adopted its magnitudes
from \citet{Carney-1993}. There are no RR Lyr variables found in NGC 6637.
The results, listed in Table \ref{fig:CMD}, were plotted as horizontal lines;
filled lines for $M_V$(RR) and dashed lines for $M_V$(RHB).
The sizes of error in these values are also calculated
as the combinations of the random scatter and the error of the reddening.
The contribution of the reddening error is large in these optical values.
We found that $M_V$(RR) and $M_V$(RHB), respectively, are common
between the four clusters, and $M_V$(RR) is brighter than $M_V$(RHB)
by 0.3--0.4 mag.

\begin{table}[!b]
\caption{Estimates of the distance moduli (DM) and absolute magnitudes of
the horizontal branch features. RHB and RR indicate red horizontal branch
and RR Lyr variables. See the text for the definitions.
\label{tab:result}}
\smallskip
\begin{center}
{
\begin{tabular}{cccccc}
\tableline
\noalign{\smallskip}
Cluster  & $E_{B-V}$ & $N_{\rm Mira}$ & DM & $M_V$(RHB) & $M_V$(RR) \\
\noalign{\smallskip}
\tableline
\noalign{\smallskip}
47 Tuc   & 0.04 & 5 & 13.15$\pm 0.01$ & 0.79$\pm 0.03$  & 0.46$\pm 0.03$ \\
NGC 6388 & 0.37 & 5 & 15.44$\pm 0.08$ & 0.66$\pm 0.24$  & 0.26$\pm 0.24$ \\
NGC 6441 & 0.47 & 8 & 15.65$\pm 0.05$ & 0.71$\pm 0.29$  & 0.43$\pm 0.29$ \\
NGC 6637 & 0.16 & 2 & 14.71$\pm 0.06$ & 0.77$\pm 0.12$  & ---   \\
\noalign{\smallskip}
\tableline
\noalign{\smallskip}
\end{tabular}
}
\end{center}
\end{table}

\section{Discussion}

In the previous section, we found that RR Lyr variables are brighter
than the red part of the HB in 47 Tuc, NGC 6388 and 6441. 
Traditionally, it is considered that metal-rich RR Lyr variables are
fainter than metal-poor ones \citep{Caputo-2000}.
In addition, \citet{Clementini-2005} conducted spectroscopic measurements
of the metallicities of RR Lyr variables in NGC 6441,
and confirmed that the variables are as metal-rich as other cluster stars.
However, $M_V$(RR) listed in Table \ref{tab:result} corresponds to
the magnitudes of RR Lyr variables with [Fe/H] $\sim -$2~dex.
It is known that the periods
of RR Lyr variables in NGC 6388 and 6441 are longer than
what expected for relatively metal-rich RR Lyr variables
(see the discussions in \citealt{Pritzl-2003}).
This feature implies that they are overluminous if based on
traditional pulsational theories, and our result confirms
that they are actually as bright as metal-poor RR Lyr variables.
The one RR Lyr variable, V9, in 47 Tuc is considered to
be a physical member of the cluster\citep{Carney-1993}, and
it is also brighter than the red part of the HB.
Its nature is quite uncertain. It may be an object which has already evolved
from the zero-age HB. Although there has been no investigations
which connect this object to the peculiar RR Lyr variables in NGC 6388 and
6441, it is interesting to study this anomalous variable in
the typical metal-rich cluster 47 Tuc.

Recently, \citet{Catelan-2006} reported a remarkable CMD
of NGC 6388 in which the turn-off of the main sequence is clearly visible.
They showed that the CMD agrees well with that of 47 Tuc
in many features, such as the positions and shapes of the turn-off,
RGB and etc, except the the blue HB. 
They claimed that the peculiar blue HB and RR Lyr variables
are overluminous, and presented its interesting implication on their natures.
However, they compared the CMDs of both clusters after they added
the shift so as to register the red parts of the HB
of the two clusters in the same position on the CMDs. On the other hand,
our result was obtained in an independent way with Mira variables
to shift the CMDs onto the absolute magnitude scale.

\section{Summary}

We presented the result of our near-infrared observations for
Mira variables in NGC 6388 and 6441,
which have peculiar HB morphology.
We obtained the periods and mean magnitudes for five and
nine Mira variables in the clusters, respectively,
although one in NGC 6441 is too bright to be a member of the cluster.
The data for the other objects were well fitted by
the period-luminosity relation in each cluster.
The distances of both clusters were then derived by adopting
the relation of Mira variables in the LMC as the reference relation at
the distance modulus of 18.5 mag. Using the derived distances,
we placed the HST optical data on the CMD
on the absolute scale, and found that their red HB stars
have similar luminosity with those of normal metal-rich clusters as 47 Tuc
while RR Lyr variables and blue HB stars are overluminous.

\acknowledgements The author acknowledges Drs. M. Feast, J. Menzies,
and P. Whitelock for helpful discussions.
Thanks are also given to Drs. Y. Nakada, T. Tanab\'e, M. Matsuura,
Ms. H. Fukushi and Mr. N. Matsumoto for their helps to improve
my talk presentation.
The author is financially supported by
the Japan Society for the Promotion of Science (JSPS) for Young Scientists.


\end{document}